\documentclass[pra,aps,showpacs,twocolumn]{revtex4} 
\usepackage{amsmath, amssymb, mathrsfs, graphicx, braket,verbatim}

\begin{document}

\bibliographystyle{unsrt}

\title{Ladder-type electromagnetically induced transparency using nanofiber-guided light in a warm atomic vapor}
\author{D.E. Jones}
\author{J.D. Franson}
\author{T.B. Pittman}
\affiliation{Physics Department, University of Maryland Baltimore
County, Baltimore, MD 21250}

\begin{abstract}

We demonstrate ladder-type electromagnetically induced transparency (EIT) using an optical nanofiber suspended in a warm rubidium vapor.  
The signal and control fields are both guided along the nanofiber, which enables strong nonlinear interactions with the surrounding atoms at relatively low powers. 
Transit-time broadening is found to be a significant EIT decoherence mechanism in this tightly-confined waveguiding geometry. 
Nonetheless, we observe significant EIT and controlled polarization rotation using control-field powers of only a few $\mu$W in this relatively robust warm-atom nanofiber system.

\end{abstract}

\pacs{42.50.Gy, 42.65.-k, 42.81.Qb, 42.62.Fi}

\maketitle

\section{Introduction}
\label{sec:introduction}

Systems allowing controllable photon-atom interactions are becoming increasingly important for quantum information applications \cite{kimble08}.  One such platform involves the interaction of the tightly confined evanescent mode of an optical nanofiber with surrounding atoms \cite{spillane08, garciafernandez11}. Because nanofibers are typically formed in the waist of tapered optical fibers (TOF's), they can be easily connected to standard fiber components with very low loss \cite{tong04}. This allows one to envision a fully fiber-based quantum network with nanofiber ``atom access points'' that can be used, for example, for quantum repeater or quantum memory stations \cite{sangouard11}.

As a very promising step in that direction, coherent storage of nanofiber-guided light pulses has recently been demonstrated using $\Lambda$-type electromagnetically induced transparency (EIT) in a cold atomic cloud \cite{gouraud15} and a trapped atomic ensemble \cite{sayrin15}. In these two systems motional effects are minimized, resulting in spectrally narrow ($\sim$kHz) EIT windows and long ($\sim\mu$s) storage times. In addition, when both the control and signal fields were guided by the nanofiber, EIT could be observed with remarkably low power ($\sim$pW) control fields \cite{sayrin15}.

In this paper we  demonstrate related nanofiber-based EIT effects in a system that differs in two primary ways: (1) we use a warm atomic vapor surrounding the nanofiber, and (2) we use ladder-type EIT  \cite{geabanacloche95} with counter-propagating control and signal fields in the nanofiber.

\begin{figure}[t]
\includegraphics[width=3in]{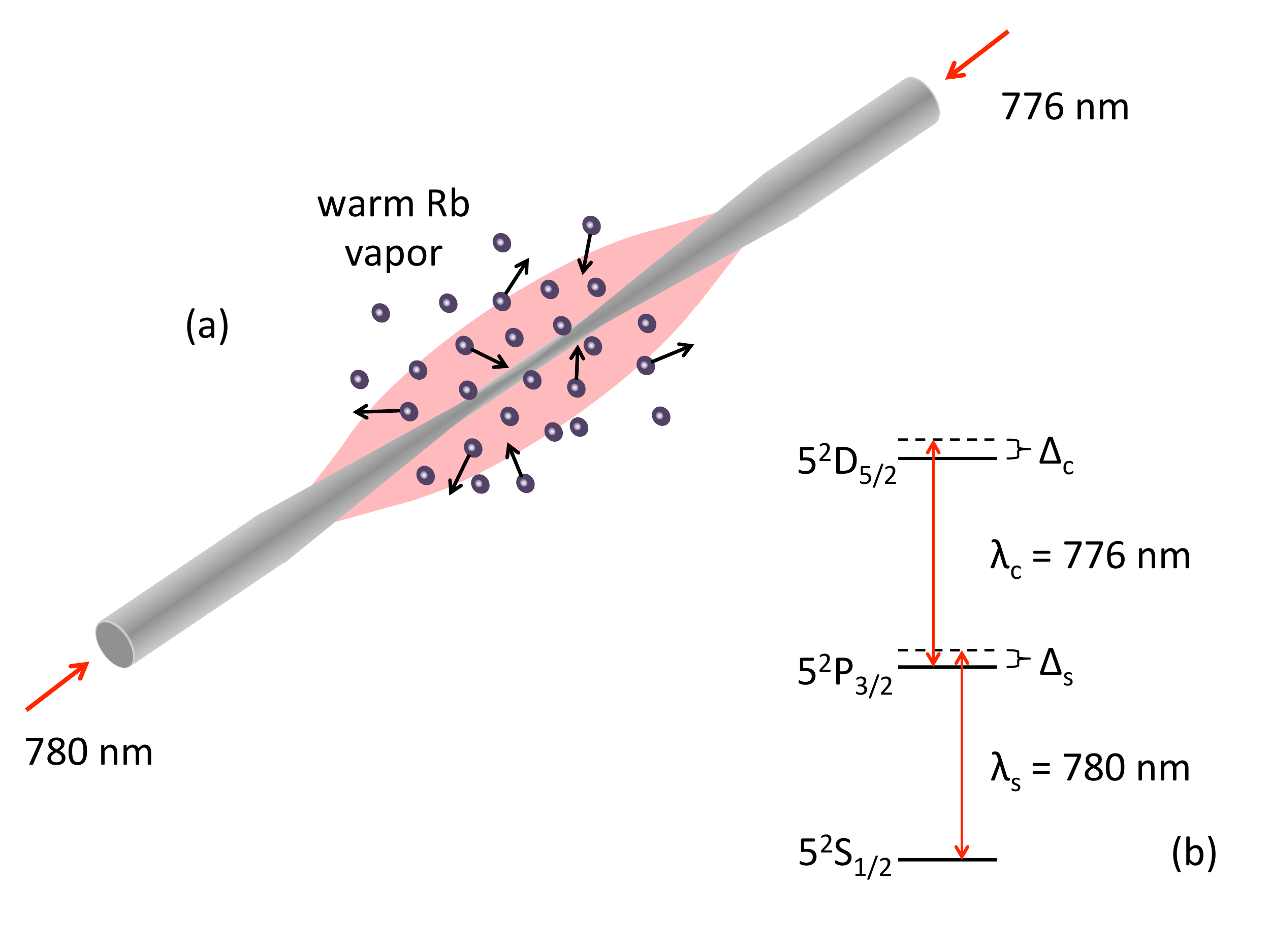}
\caption{(Color online) Overview of the system: (a) An optical nanofiber surrounded by warm rubidium vapor. (b) 3-level ladder-type EIT system using the $5S_{1/2}\rightarrow5P_{3/2}\rightarrow5D_{5/2}$ two-photon transition in rubidium. As depicted in (a), the signal (lower) beam at 780 nm and control (upper) beam at 776 nm counterpropagate through the nanofiber. The small diameter of the evanescent optical mode results in thermal atoms quickly moving through the control and signal fields on a timescale of a few ns.}
\label{fig:fig1}
\end{figure}

An overview of the specific system is shown in Fig. \ref{fig:fig1}.  A sub-wavelength diameter nanofiber formed in the waist of a standard TOF is surrounded by a warm vapor of rubidium atoms. We use the  $5S_{1/2}\rightarrow5P_{3/2}\rightarrow5D_{5/2}$ two-photon ladder transition with the signal (lower) field at 780 nm and the control (upper) field at 776 nm. EIT is first observed by examining strong modification of the 780 nm signal absorption spectrum with the application of a resonant 776 nm control field.   As an example of the utility of this warm-atom nanofiber EIT system, we then demonstrate coherent control of the polarization of the signal field using the method developed by Wielandy and Gaeta (WG) \cite{wielandy98a}.   Here, the 776 nm control beam is used to induce birefringence in the atomic vapor that causes a rotation of the polarization of the 780 nm signal beam.
 
In analogy to EIT with free-space beams in warm vs. cold atomic ensembles \cite{novikova12}, the warm-atom nanofiber system used here is easier to implement than the cold-atom nanofiber system, but the EIT effects are significantly reduced due to thermal motion of the atoms.  Although Doppler broadening of the EIT window is largely cancelled in our system, significant transit-time broadening remains due to the very short time ($\sim$ ns) that the thermal atoms spend traversing the small evanescent mode ($\sim 1 \,\mu$m diameter) of the nanofiber \cite{hendrickson10}. 

Despite this significant broadening mechanism, we are able to see clear evidence of EIT with control field powers of only a few $\mu$W's in this system.   In addition, by detuning the resonant control field to optimize the WG rotation effect, we observe $\sim$2\% transmission of the probe field through a crossed analyzer with a control power of only 20 $\mu$W. The observations of these two effects (EIT and controlled polarization rotation) in the warm-atom nanofiber system are the main results of this paper

\section{Experiment}
\label{sec:experiment}

A TOF was pulled from standard single-mode fiber using the flamebrush technique \cite{birks92} to realize a nanofiber with a central waist diameter of $\sim$300 nm and a sub-500 nm diameter over a length of 8 mm. The TOF was installed in a vacuum system using a specialized nanofiber heating unit designed to minimize the accumulation of rubidium on the nanofiber surface \cite{lai13}.  The atomic density was controlled by heating a metallic rubidium sample in the vacuum system.  In this warm-atom nanofiber system, we typically achieved optical depths (OD's) of $\sim$3 for the transition of interest (see dashed box region in Figure 2) at rubidium temperatures of $\sim 85^{\circ}$C.

The signal (780 nm) and control (776 nm) fields were generated by two independent fiber-coupled narrowband tunable diode lasers (linewidths $\sim$300 kHz) that were sent in counter-propagating directions through the nanofiber in order to achieve (nearly) Doppler-free two-photon effects \cite{hendrickson10}. The counter-propagating signal and control lasers were also sent into a standard free-space rubidium vapor cell setup that could be used to calibrate the system and simultaneously compare nanofiber-based EIT effects with the same effects observed in a conventional free-space beam geometry \cite{wielandy98a, wielandy98b}. 

The 780 nm and 776 nm output signals were isolated using narrowband interference filters; the combination of these filters and the counter-propagating geometry resulted in good signal-to-noise ratios for  low-power ($\sim$ nW) measurements with conventional amplified photodiodes. Additional details on the basic warm-atom nanofiber vacuum system and overall experimental setup can be found in \cite{jones14}.

\section{Electromagnetically Induced Transparency}
\label{sec:overview}

Experimental results demonstrating the ability of the control beam to modify the transmission of a low power (10 nW) signal beam through the warm-atom nanofiber system are shown in Figure 2. First, the upper trace shows the 780 nm signal transmission spectrum without the 776 nm control beam applied. The interaction of the evanescent mode of the nanofiber with the surrounding rubidium vapor is evidenced by the four Doppler-broadened absorption dips which are due to the two ground-state hyperfine levels for each isotope ($^{85}$Rb and $^{87}$Rb) in the natural rubidium vapor \cite{steck}. Next, the lower trace shows the same transmission spectrum, but with a 7 $\mu$W  776 nm control field applied. Here, EIT windows are evident near the centers of each of the four absorption dips. 

In Figure 2, the 780 nm signal field detuning $\Delta_{s}$ is defined relative to the  $^{85}$Rb $5S_{1/2}\, (F=2) \rightarrow 5P_{3/2}\, (F'=1,2,3)$ transition at 384.232 THz.  In addition, the 776 nm control field detuning $\Delta_{c}$ is defined relative to the  $^{85}$Rb $5P_{3/2}\, (F'=1,2,3) \rightarrow 5D_{5/2}\, (F''=0\!-\!4)$ transition at 386.340 THz.  For the data shown in Figure 2(b), the control field detuning was held fixed at $\Delta_{c}=0$.  The transparency window at $\Delta_{s} = 0$, for example, has $\sim$20\% transmission and a width of $\sim$200 MHz.

\begin{figure}[t]
\includegraphics[width=3in]{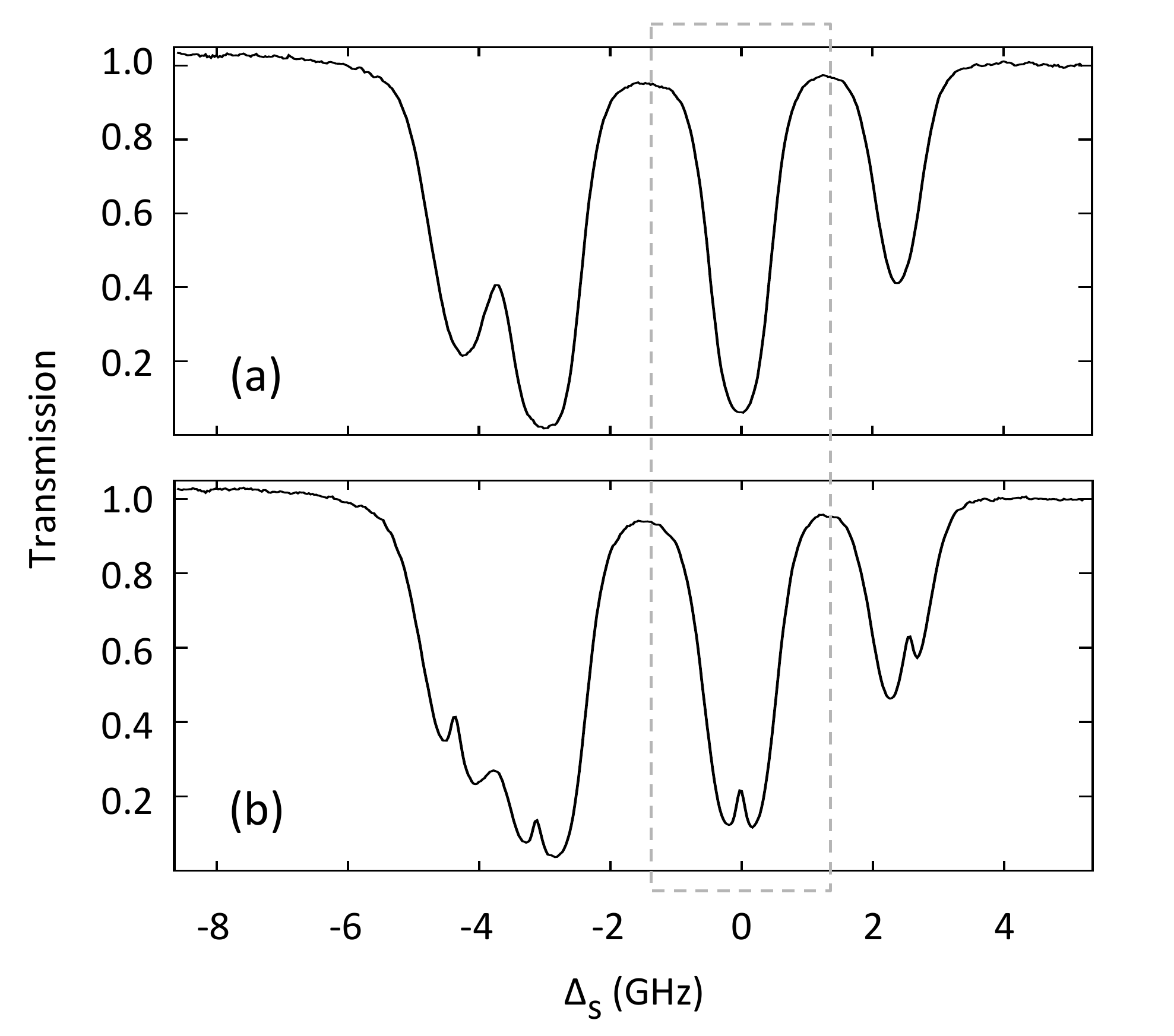}
\caption{Experimental evidence of EIT in the warm-rubidium nanofiber system. The scans show transmission of a 780 nm signal beam (a) without a 776 nm control beam, and (b) with a 7 $\mu$W resonant 776 nm control beam ($\Delta_{c}=0$).   With the application of the control beam, narrow transparency windows are seen within the centers of each of the four Doppler-broadened absorption dips.  The signal frequency detuning $\Delta_{s}$ is defined relative to the $^{85}$Rb $5S_{1/2}\, (F=2) \rightarrow 5P_{3/2}\, (F'=1,2,3)$ transition.  The transmission is defined relative to the overall TOF system transmission ($\sim$35\%)  far from resonance.
The dashed box denotes the region of interest for the remainder of the paper.}
\label{fig:fig2}
\end{figure}

In order to highlight the role of transit-time broadening in this particular EIT effect, Figure 3 shows a calculation of the imaginary part of the susceptibility using the semi-classical model developed in \cite{geabanacloche95}. The plot shows the expected 780 nm signal absorption in the vicinity of the transition at $\Delta_{s}=0$ (i.e., in the dashed-box region of Figure 2). The green curve in Figure 3 shows the warm-atom nanofiber system considered here, where both Doppler and transit-time effects are included.  In contrast, the red curve shows the more familiar case of large-diameter free-space beams in a warm rubidium vapor cell, where Doppler broadening is also significant but transit-time effects can be ignored.

\begin{figure}[t]
\includegraphics[width=3in]{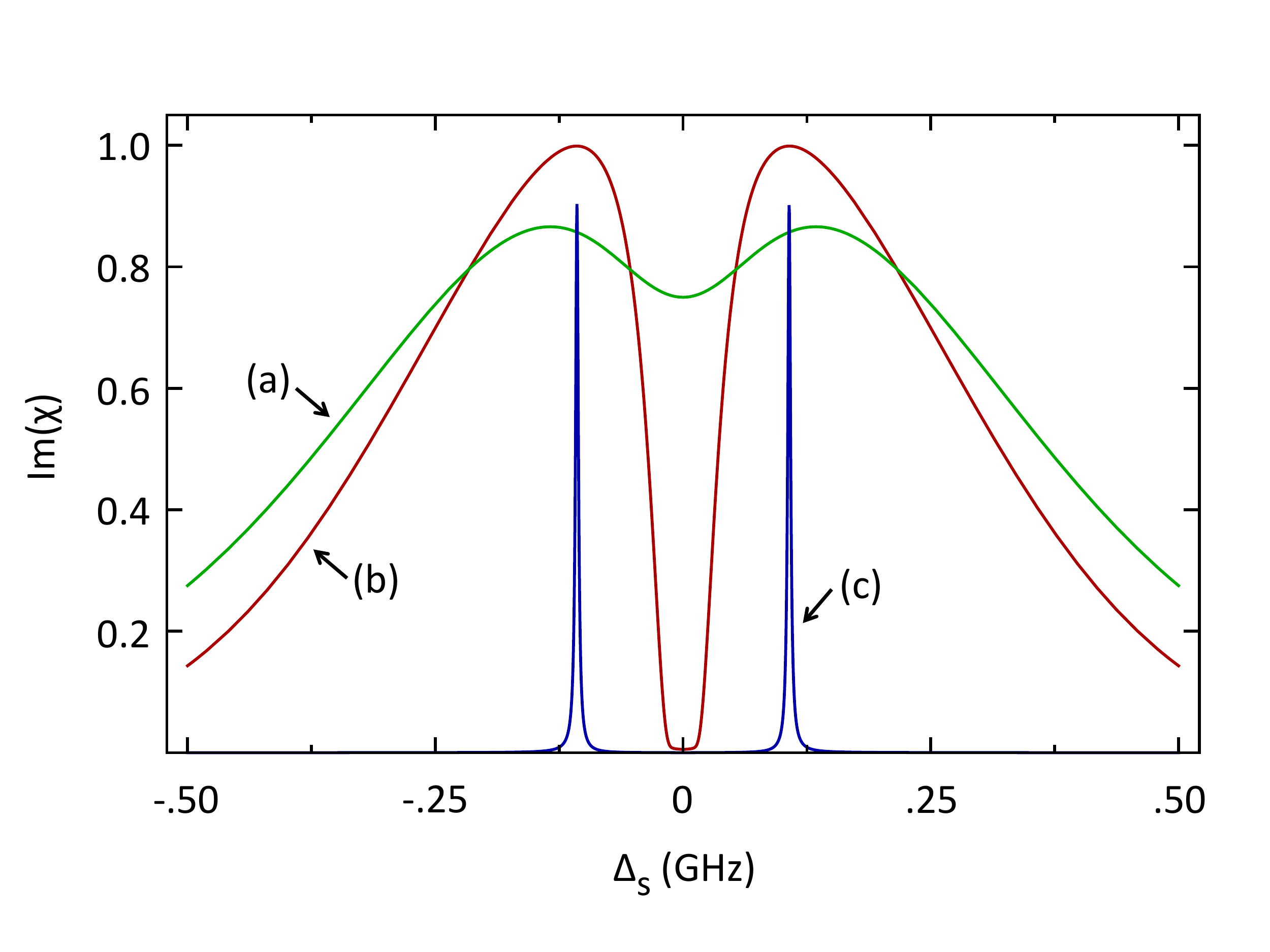}
\caption{(Color online) Theoretical calculation of the imaginary part of the normalized susceptibility for the 780 nm signal beam with the application of a resonant 776 nm control beam ($\Delta_{c}=0$).  The green curve (a) corresponds to  the warm-rubidium nanofiber system considered here, where both Doppler broadening and transit-time broadening are significant. The control-field Rabi frequency was chosen to simulate the experimentally observed EIT in Figure 2. For comparison, the red curve (b) shows the same Rabi frequency for the case of a free-space rubidium vapor cell, where transit-time broadening is neglected and, consequently, the transparency is much deeper.  For reference, the blue curve (c) corresponds to the case of cold-atom EIT, where all motional effects are neglected.}
\label{fig:fig3}
\end{figure}

For the models used in Figure 3, the Doppler width ($\sim$570 MHz) and transit time broadening ($\sim$100 MHz) were determined by the typical rubidium temperature ($85^{\circ}$) used in our experiments.  For simplicity, additional broadening due to atomic collisions (and collisions with the nanofiber itself) were neglected.  The control-field Rabi frequency (214 MHz) was chosen so that the green curve closely matched the experimentally observed nanofiber EIT effect shown in Figure 2.  This Rabi frequency is consistent with the expected value for a $\mu$W-level control field in a typical nanofiber mode geometry \cite{tong04}.

It can be seen in Figure 3 that the Rabi frequency (ie. control field intensity) needed to produce a $\sim$20\% transparency window in the warm-atom nanofiber system is enough to produce a complete 100\% transparency window in a typical free-space vapor cell system.  The large difference in these transparency values shows the significance of transit-time broadening in the warm-atom nanofiber system.  For further comparison, the blue curve in Figure 3 shows the same model and Rabi frequency applied to the case of cold atoms, where all motional effects (Doppler and transit-time) can be neglected \cite{gouraud15,sayrin15}.

Figure 4 shows additional experimental scans of the EIT window at $\Delta_{s} = 0$ with control field powers ranging from 200 nW to 45 $\mu$W.  The onset of EIT becomes obvious at control powers as low as a few $\mu$W in this warm-atom nanofiber system.  As the control power is increased, the transparency increases and begins to experience power broadening. The rapid broadening of the overall absorption dip and the transparency window at the highest control power shown is indicative of significant contributions due to Autler-Townes Splitting (ATS) \cite{anisimov11, peng14}.  Indeed, ATS can occur at extremely low control powers due to the nanofiber's tightly confined mode geometry, especially for the case of a cold-atom nanofiber system with negligible Doppler broadening \cite{kumar15, tan14}.  The transition from EIT-dominated to ATS-dominated transparency at higher control powers in a warm-atom nanofiber system also displays interesting characteristics \cite{jones15}.

\begin{figure}[t]
\includegraphics[width=3in]{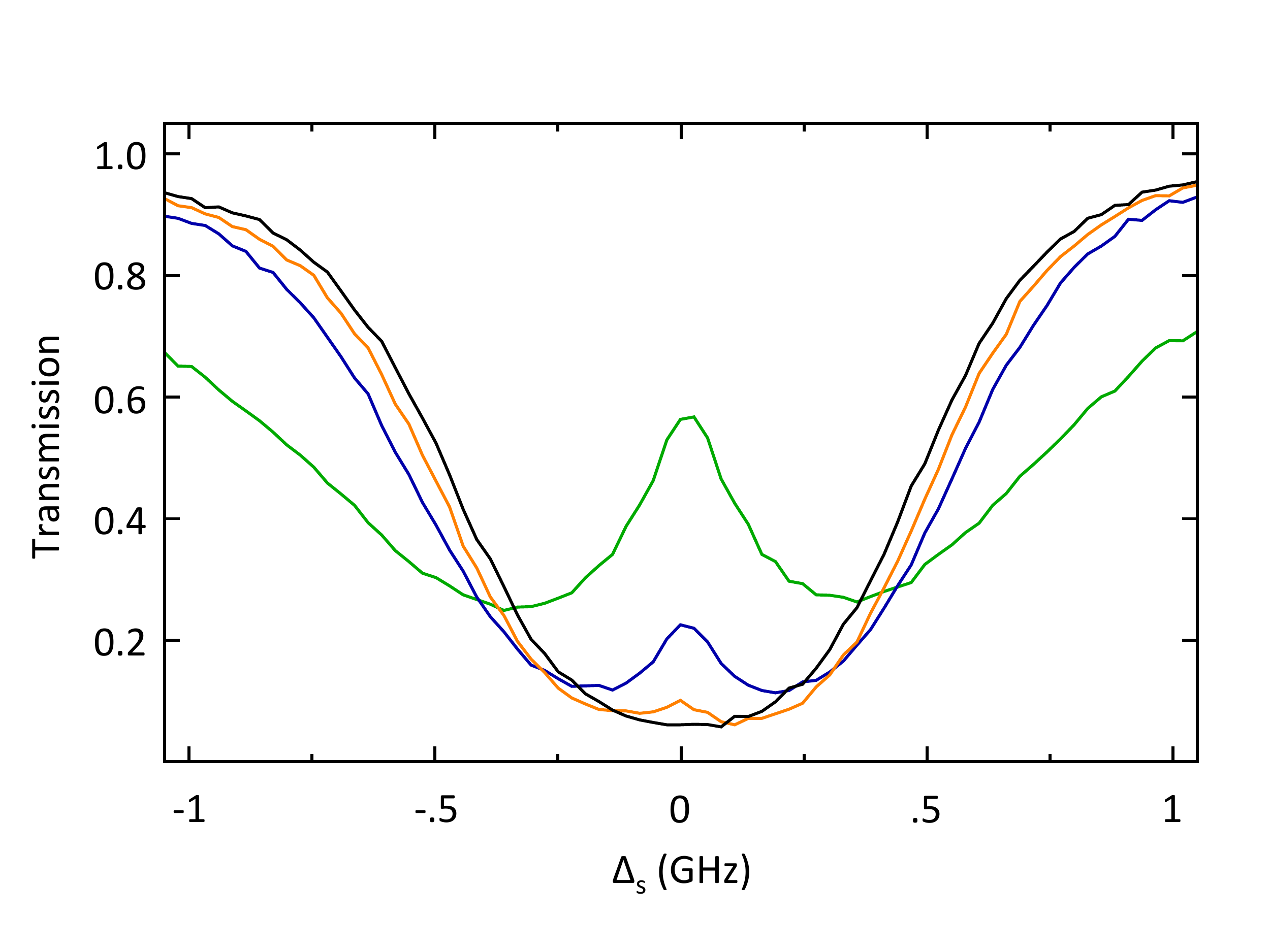}
\caption{(Color online) Experimental observation of EIT as a function of resonant control field power for the warm-atom nanofiber system. The plot shows increasing signal field transmission near $\Delta_{s} = 0$ (the $^{85}$Rb $5S_{1/2}\, (F=2) \rightarrow 5P_{3/2}\, (F'=1,2,3)$ transition) for increasing control field powers of 200 nW (black), 2 $\mu$W (orange), 7 $\mu$W (blue) and 45 $\mu$W (green).}
\label{fig:fig4}
\end{figure}

\section{Controlled Polarization Rotation}

The controlled absorption experienced by the signal field in Figs. 2-4 is accompanied by controlled phase shifts. Because the magnitude of these phase shifts is different for various transitions among the magnetic sub-levels, the 776 nm control beam can be used to induce a birefringence in the atomic vapor that causes a change in the polarization of the 780 nm signal beam \cite{wielandy98a}. 

Figure 5(a) shows one example of this effect using the specific $5S_{1/2} (F=2) \rightarrow 5P_{3/2} (F'=3) \rightarrow 5D_{5/2} (F''=4)$ two-photon transition in  $^{85}$Rb. For simplicity, we assume that the initial population is all in the  $m_{F} =0$ magnetic sub-level of the ground state.   The control field is chosen to be $\sigma^{+}$ polarized. When the signal field is chosen to be linearly polarized (ie. a superposition of $\sigma^{+}$  and $\sigma^{-}$), selection rules give two different EIT pathways indicated by the red and green arrows. Because the two-photon transition amplitude for the green path ($\sigma^{+}$:$\sigma^{+}$) is significantly stronger than the red path ($\sigma^{-}$:$\sigma^{+}$) \cite{mcgloin00}, this results in a control-induced circular birefringence experienced by the signal beam \cite{cho05}.

\begin{figure}[t]
\includegraphics[width=3.25in]{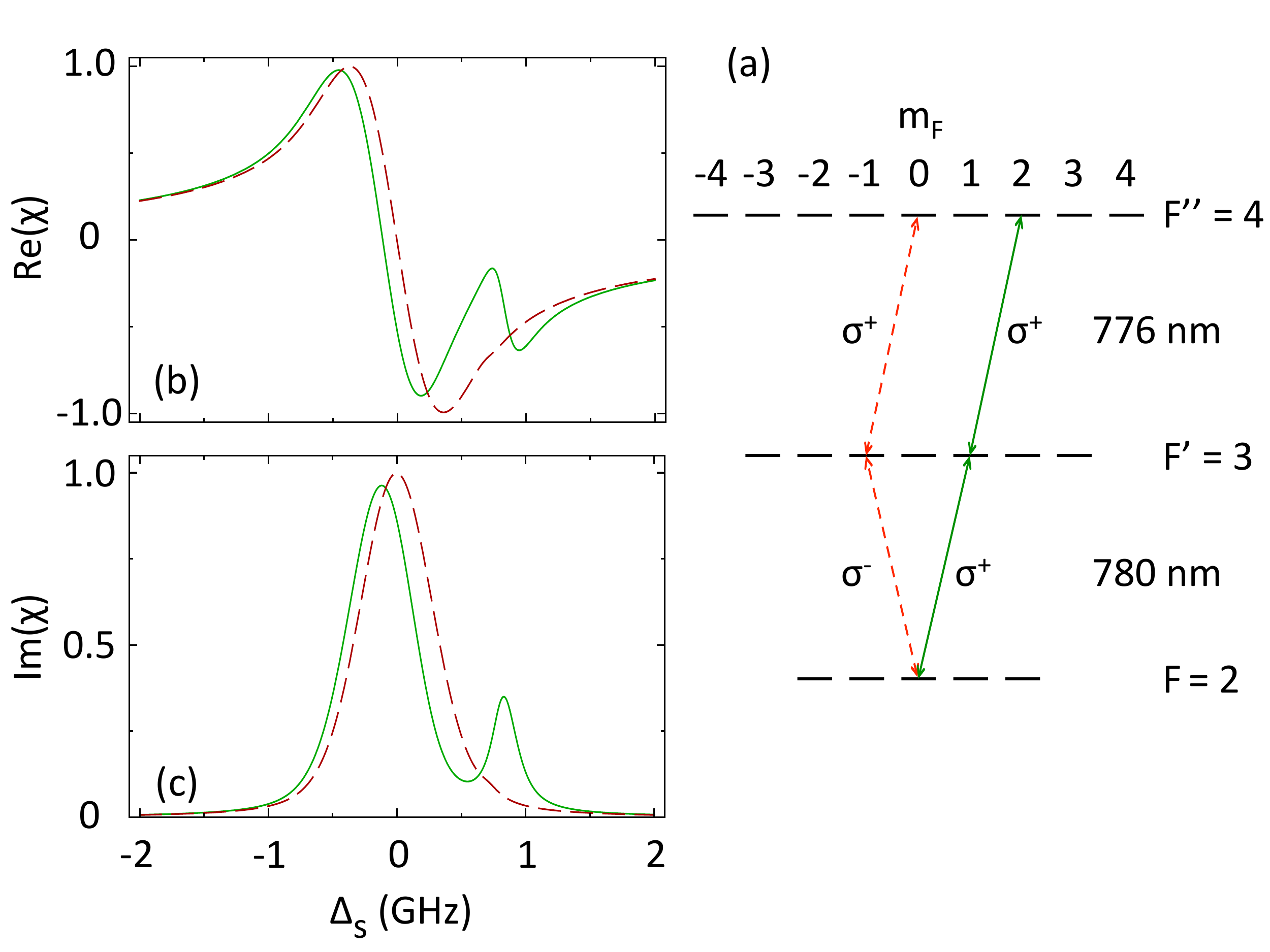}
\caption{(Color online) A simplified example of controlled polarization rotation using a $\sigma^{+}$ polarized 776 nm control field and a linearly polarized 780 nm signal field.  (a) shows the two possible EIT pathways ($\sigma^{-}:\sigma^{+}$ [dashed red] and  $\sigma^{+}:\sigma^{+}$ [solid green] ), among the various magnetic sublevels for the case when all of the initial population starts in the $m_{F}=0$ sublevel of the ground state. The different strengths of these two pathways induces a circular birefringence for the signal field.  (b) and (c) show calculations of the normalized signal-field susceptibility for a control-field detuning fixed at $\Delta_{c}=700$ MHz.  }
\label{fig:fig5}
\end{figure}

A calculation of the real and imaginary parts of the signal-field susceptibility for these two EIT pathways is shown in Figures 5(b) and 5(c). Once again, the semi-classical model of \cite{geabanacloche95} was used, with the relevant Doppler broadening and transit time effects of the warm-atom nanofiber system included.  A control-field detuning $\Delta_{c}=700$ MHz was chosen to optimize the trade-off between maximizing the difference in  phase shifts experienced by the $\sigma^{+}$ and $\sigma^{-}$ components of the signal field while simultaneously minimizing the loss of each one \cite{wielandy98a}. The control field Rabi frequency was chosen to correspond to a typical control power used in our experiments. For this example it can be seen that a large signal-field polarization rotation should be expected at a signal detuning in the vicinity of $\Delta_{s}\sim 700$ MHz.

\begin{figure}[t]
\includegraphics[width=3.25in]{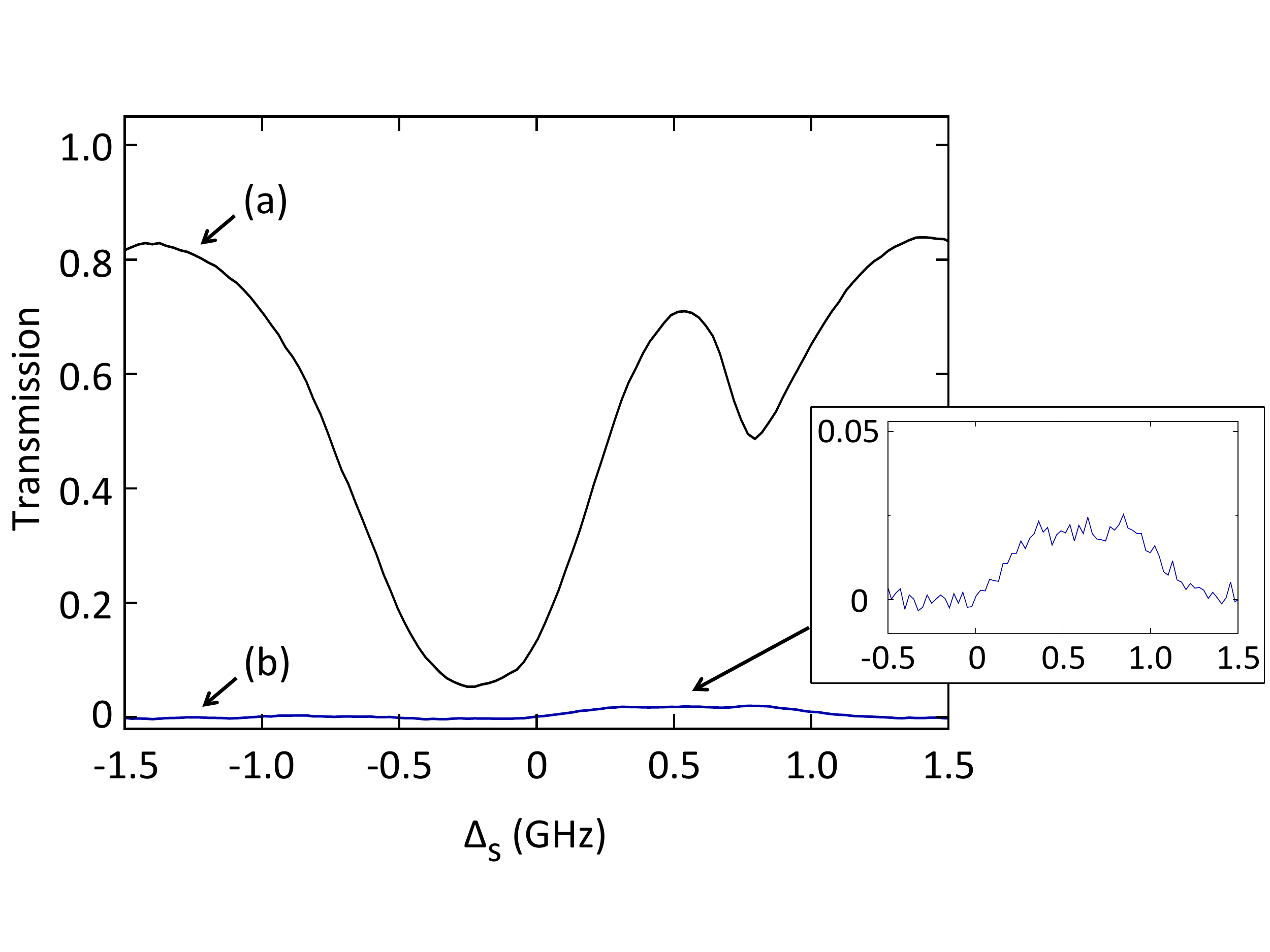}
\caption{(Color online) Experimental demonstration of controlled polarization rotation using the warm-atom nanofiber system. The black curve (a) shows the transmission of the signal beam in the presence of a detuned control beam ($\Delta_{c} \sim 700$ MHz) with a power of 20 $\mu$W; the data is analogous to the EIT data of Figure 4, but with the transparency window shifted from the center due to the non-zero control detuning. The blue curve (b) shows the same situation but with a crossed polarizer on the output (here the signal transmission is normalized to transmission through a parallel polarizer). The inset shows a ``zoom-in'' on the same data highlighting the controlled polarization rotation. }
\label{fig:fig6}
\end{figure}

Figure 6 shows an experimental demonstration of this type of controlled polarization rotation using the warm-atom nanofiber system. For this data, the power of the $\sigma^{+}$ polarized control beam was 20 $\mu$W, and its detuning was fixed at $\Delta_{c} \sim 700$ MHz.  The signal field was linearly polarized and had a power of 90 nW.  First, the black curve in Figure 6 shows the transmission of the signal beam through the system as a function of $\Delta_{s}$ without any polarizers on the output. The presence of the control beam results in the same type of EIT shown in Figure 4, but with the transparency window shifted from the center of the Doppler-broadened absorption dip because $\Delta_{c} \sim 700$ MHz.

Next, the blue curve in Figure 6 shows the transmission of the signal beam through the system with a crossed polarizer on the output. The significant transmission ($\sim$2\%) of the signal field through the crossed polarizer near $\Delta_{s}\sim700$ MHz demonstrates a controlled-polarization rotation of roughly $8^{\circ}$ based on the arguments of Figure 5. 

It is important to note that the magnitude of the experimentally-observed signal transmission through the crossed  polarizer in Figure 6 is significantly smaller than what would be predicted from the simplified model of Figure 5 due to two main factors. First, the initial population in the warm rubidium vapor was distributed over all of the $m_{F}$ sub-levels of the $F=2$ ground state, and the closely spaced $F'$ intermediate states and $F''$ upper states were within the Doppler width of the system. Consequently, there were many more two-photon transitions involved than the two shown in Figure 5.

Second, and more importantly, the polarization state of the nanofiber guided light interacting with the atoms is highly complex and difficult to control \cite{kien04}. This was further complicated by varying stress-induced birefringence along the length of nanofiber due to ``slack'' in the TOF which results from the particular mounting procedure we used \cite{lai13}. Consequently, the polarizations of the control and signal field were not the ideal states shown in Figure 5 over the length of the interaction region; in practice, we simply optimized them by maximizing the controlled polarization rotation signal itself.

Nonetheless, the ability to significantly alter the polarization of the signal field using control powers of only 10's of $\mu$W highlights the ability to perform ultralow-power nonlinear optics in the warm-atom nanofiber system. For very rough comparison, EIT-type polarization rotation experiments in standard free-space vapor cell systems typically use control field powers of 10's to 100's of mW \cite{wielandy98a,li06,krishnamurthy14}. In both systems, higher control powers generally result in larger polarization changes, but the exact birefringence experienced by the signal field is a complicated function of the control field power \cite{wielandy98a}.

\section{Summary and Discussion}

Optical nanofibers surrounded by atomic ensembles have become a promising platform for controlled photon-atom interactions \cite{spillane08, garciafernandez11, schroder12, pittman13, lee15}. Strong nonlinearities with low-power fields can be realized in this system due to the propagation of nanofiber-guided evanescent modes with very small cross-sectional areas over relatively long distances through the surrounding atoms.  In the context of EIT, this allows the observation of large transparencies with control-field powers that are several orders of magnitude smaller than those used in comparable free-space beam platforms.

In this paper we experimentally observed (1) ladder-type EIT \cite{geabanacloche95} and (2) coherent control of the signal field polarization \cite{wielandy98a} with control-field powers of only a few $\mu$W using a warm-atom nanofiber system.  The dominant EIT decoherence mechanism was found to be transit time broadening due to the short time the atoms spend moving through the small evanescent mode of the nanofiber. This problem can essentially be avoided using cold-atom or trapped-atom nanofiber EIT systems, allowing even lower control-field powers and longer EIT-storage times at the cost of more complex experimental systems \cite{gouraud15, sayrin15, kumar15} .  Nonetheless, the ability to perform $\mu$W-level all-optical control using the relatively simple and robust system demonstrated here may be useful for practical low-power all-optical applications.

This work was supported by the NSF under grant No. 1402708 and by DARPA DSO under grant No. W31P4Q-12-1-0015.

Note:  During the preparation of this manuscript, we became aware of similar work done in a cold-atom nanofiber system, in which all-optical switching was demonstrated \cite{kumar15b}.




\begin{thebibliography}{50}

\bibitem{kimble08} H.J. Kimble, Nature {\bf 453}, 1023 (2008). 

\bibitem{spillane08} S.M. Spillane, G.S. Pati, K. Salit, M. Hall, P. Kumar, R.G. Beausoleil, and M.S. Shahriar, Phys. Rev. Lett. {\bf 100}, 233602 (2008).

\bibitem{garciafernandez11} R. Garcia-Fernandez, W. Alt, F. Bruse, C. Dan, K. Karapetyan, O. Rehband, A. Stiebeiner, U. Wiedemann, D. Meschede, and A. Rauschenbeutel, Appl. Phys. B {\bf 105}, 3 (2011).

\bibitem{tong04} L. Tong, J. Lou, and E. Mazur, Opt. Exp. {\bf 12}, 1025 (2004).

\bibitem{sangouard11} N. Sangouard, C. Simon, H. de Riedmatten, and N. Gisin, Rev. Mod. Phys. {\bf 83}, 33 (2011).

\bibitem{gouraud15} B. Gouraud, D. Maxein, A. Nicolas, O. Morin, and J. Laurat, Phys. Rev. Lett. {\bf 114}, 180503 (2015).

\bibitem{sayrin15} C. Sayrin, C. Clausen, B. Albrecht, P. Schneeweiss, and A. Rauschenbeutel, Optica {\bf 2}, 353 (2015).

\bibitem{geabanacloche95} J. Gea-Banacloche, Y.-Q. Li, S.-Z. Jin, and M. Xiao, Phys. Rev. A {\bf 51}, 576 (1995).

\bibitem{wielandy98a} S. Wielandy and A.L. Gaeta, Phys. Rev. Lett. {\bf 81}, 3359 (1998).

\bibitem{novikova12} I. Novikova, R.L. Walsworth, and Y. Xiao, Laser \&  Photon. Rev. {\bf 6}, 333 (2012).

\bibitem{hendrickson10} S.M. Hendrickson, M.M. Lai, T.B. Pittman, and J.D. Franson, Phys. Rev. Lett. {\bf 105}, 173602 (2010).

\bibitem{birks92} T. Birks and Y. Li, J. Lightwave Technol. {\bf 10}, 432-438 (1992).

\bibitem{lai13} M.M. Lai, J.D. Franson, and T.B. Pittman, Appl. Opt. {\bf 52} 2595 (2013).

\bibitem{wielandy98b} S. Wielandy and A.L. Gaeta, Phys. Rev. A {\bf 58}, 2500 (1998).

\bibitem{jones14} D.E. Jones, J.D. Franson, and T.B. Pittman, J. Opt. Soc. Am. B {\bf 31}, 1997 (2014).

\bibitem{steck} D. A. Steck, ``Rubidium 85 D Line Data," 2012, http://steck.us/alkalidata/.

\bibitem{anisimov11} P.M. Anisimov, J.P. Dowling, and B.C. Sanders,  Phys. Rev. Lett {\bf 107}, 163604 (2011).

\bibitem{peng14} B. Peng, S.K. Ozdemir, W. Chen, F. Nori, and L. Yang, Nat. Commun. {\bf 5}, 5082 (2014)

\bibitem{kumar15} R. Kumar, V. Gokhroo, K. Deasy, and S.N. Chormaic, Phys. Rev. A {\bf 91}, 053842 (2015).

\bibitem{tan14} The control-field Rabi frequencies needed for significant ATS in Doppler broadened systems (such as ours) are higher than those needed in Doppler-free systems:  C. Tan and G. Huang, J. Opt. Soc. Am. B {\bf 31}, 704 (2014).

\bibitem{jones15} D.E. Jones, J.D. Franson, and T.B. Pittman, {\em in preparation} (2015).

\bibitem{mcgloin00} D. McGloin, M.H. Dunn, and D.J. Fulton. Phys Rev. A {\bf 62}, 053802 (2000)

\bibitem{cho05} D. Cho, J.M. Choi, J.M. Kim, and Q-H. Park, Phys. Rev. A {\bf 72}, 023821 (2005).

\bibitem{kien04} F.L. Kien, J.Q. Liang, K. Hakuta, and V.I. Balykin, Opt. Commun. {\bf 242}, 445 (2004).

\bibitem{li06}  S. Li, B. Wang, X. Yang, Y. Han, H. Wang, M. Xiao, and K.C. Peng, Phys. Rev. A {\bf 74}, 033821 (2006).

\bibitem{krishnamurthy14} S. Krishnamurthy, Y. Tu, Y. Wang, S. Tseng, and M.S. Shahriar, Opt. Exp. {\bf 22}, 28898 (2014).

\bibitem{schroder12} T. Schroder, M. Fujiwara, T. Noda, H.-Q. Zhao, O. Benson, and S. Takeuchi, Opt. Exp. {\bf 20}, 10490 (2012).

\bibitem{pittman13} T.B. Pittman, D.E. Jones, and J.D. Franson, Phys. Rev. A {\bf 88}, 053804 (2013).

\bibitem{lee15} J. Lee, J.A. Grover, J.E. Hoffman, L.A. Orozco, and S.L. Rolston, J. Phys. B: At. Mol. Opt. Phys. {\bf 48}, 165004 (2015).

\bibitem{kumar15b} R. Kumar, V. Gokhroo, and S.N. Chormaic, arXiv:1507.06451.

\end{thebibliography}
\end{document}